\documentclass[12pt]{article}
\usepackage{lineno}

\usepackage[margin=1in]{geometry}
\usepackage{amssymb,amsmath,latexsym}                        
\usepackage{graphicx}
\usepackage{cite}
\usepackage{xcolor}
\usepackage{float}

\usepackage{amsthm}
\usepackage{multicol}
\usepackage{authblk}

\usepackage[colorlinks]{hyperref}

\usepackage{adjustbox}
\usepackage{bbm}
\usepackage{mathtools}

\usepackage{graphicx}
\usepackage{subcaption}
\date{} 

\begin{document}
\title{A detailed study on the intrinsic time resolution of the future MRPC detector}
\author[1]{Fuyue Wang}
\author[1]{Dong Han}
\author[1]{Yi Wang\footnote{Corresponding author. Email: yiwang@mail.tsinghua.edu.cn.}}
\author[1]{Pengfei Lyu}
\author[1]{Yuanjing Li}
\affil[1]{\normalsize\it Tsinghua University\\Key Laboratory of Particle and Radiation Imaging, Ministry of Education, Beijing 100084, China}

\maketitle
 
\begin{abstract}
A challenging goal of 20 ps for the total time resolution of the Multi-gap Resistive Plate Chambers (MRPCs) system has been recently addressed by high energy physics experiments. In order to meet this requirement, one should have a deeper understanding of the detector physics and the factors that contribute to the intrinsic time resolution of the MRPC, which are studied in this paper. The sources of the timing uncertainties, their influences and the relationship with the detector geometry are analyzed. A comparison between the time resolutions obtained using two different simulation algorithms is presented. The obtained results are useful for the design of the timing MRPCs with an improved performance, and meanwhile offer guidance for the R\&D activities in the field.
\end{abstract}

\section{Introduction}
\label{sec:intro}
The Multi-gap Resistive Plate Chambers (MRPC) is a gaseous detector\cite{zeballos1996new} with high time resolution and high efficiency. It has been widely used as a basic component of the Time-of-Flight (ToF) systems of many high energy physics experiments. The published results of the total time resolution of the MRPCs systems used in the present or designed for the future experiments are in the range of 50$\sim$70 $\rm ps$\cite{akindinov2009final,Toia:209729,bonner2003single,yong2012prototype}. However, in the future Solenoidal Large Intensity Device (SoLID) at Jefferson Lab (JLab), the total time resolution of the MRPC system is required to be around 20 $\rm ps$\cite{solid2017solid} so as to achieve a 3$\sigma$ separation of $\pi/K$ up to the momentum of 7 $\rm GeV/c$. The MRPC system contains the detector and the read out units including the transmission lines and the electronics. The requirement of 20 ps for the system is supposed to be equivalent to around 15 ps for only the MRPC detector. In order to verify the achievability of this goal, a very detailed study of the intrinsic time resolution of MRPCs is needed before its design and construction. (Unless otherwise stated, the time resolution in the following parts of the paper means the intrinsic time resolution.)

MRPC originates from the Resistive Plate Chamber (RPC)\cite{santonico1981development} and their working mechanisms are very similar. Previous studies of the intrinsic time resolution are mainly focused on the RPC detectors, especially for the cases when signals start from only one single electron\cite{riegler2003detector}. This model can approximately explain the performance of the RPC detector with only one or two gas gaps, but is far from satisfactory for the MRPC, which has much more gas gaps (usually more than 5, or in a particular case, even more than 20\cite{an200820}). There are also some studies of the intrinsic time resolution focusing on avalanche signals developed from multiple primary electrons\cite{blanco2003resistive,riegler2009time}. The time response function of these signals is derived and its relationships with the threshold and attachment coefficient are discussed. Though these studies are not based on the MRPC detector that can achieve very high intrinsic time resolution, i.e. around 15 ps, they still provide ideas on how the problems can be formulated and solved.


This work presents a detailed study of the intrinsic time resolution of the future MRPC detectors, mainly from three aspects: the sources of the timing uncertainties in terms of the detector physics, their quantitative contributions to the time resolution, and the influence of the reconstruction algorithms. The traditional reconstruction algorithm is called the Time-over-Threshold (ToT)\cite{anghinolfi2004nino}, which sets a fixed threshold to the induced signal and discriminates the crossing time. Since almost all the previous results of the detector are obtained using ToT, a much simpler analysis of the intrinsic time resolution based on this method is introduced. The sources of the timing uncertainties were extracted and their quantitative contributions to the intrinsic time resolution were analyzed by the simulation. The impact of geometrical factors, (i.e. gap size and number of gaps), on the time resolution was investigated, with the aim to develop MRPCs with an improved time resolution. 

Recently, a new algorithm based on several sets of the neural networks has been proposed\cite{wang2018sorma} to further improve the detector resolution. Theoretical analysis of the network is quite difficult, but the intrinsic performance of this method can be estimated and explained by the simulation, which is also included in this paper.

The paper is organized as follows: Sec.\ref{sec:theory} describes the theoretical derivations of the intrinsic time resolution for MRPCs analyzed with the ToT method. Sec.\ref{sec:source} presents the sources of the timing uncertainties and their quantitative estimations for several typical MRPCs. Sec.\ref{sec:geo} shows the resolution with respect to some important geometry factors. Sec.\ref{sec:NN} describes the time resolution obtained with the neural network. Finally, Sec.\ref{sec:concl} concludes this paper.

\section{Study of the intrinsic time resolution with ToT method for $m$ primary electrons} 
\label{sec:theory}
When particles impinge on the MRPC detector, they will interact with the working gas and the ionized electron-ion pairs are created. The electrons will immediately drift toward the anode and trigger avalanche multiplication under the electric field. According to the Townsend effect\cite{townsend1900}, in the avalanche, the average number of electrons $\bar{n}(t)$ developed from one single primary electron grows approximatively in an exponential way, which is:
\begin{equation}
\begin{aligned}
\label{eq:townsend}
\bar{n}(t)=e^{(\alpha-\eta)vt}=e^{\alpha'vt}
\end{aligned}
\end{equation}
where $t$ is the drift time and $v$ is the drift velocity. $\alpha$ is the first Townsend coefficient and $\eta$ is the attachment coefficient. $\alpha'=\alpha-\eta$ is the first effective Townsend coefficient. However, considering the randomness, uncertainty exists around the exponential multiplication. Suppose there is an electron created at position $z=0$ and drifting along the $z$ axis under the electric field. The probability for it to become $n$ electrons after some drifting time $t$ is $P(n,t)$\cite{riegler2003detector}:

\begin{equation}
\label{eq:ampliexp}
P(n,t)=\left\{
\begin{aligned}
&k, &n=0 \\
&\frac{(1-k)^2}{\bar{n}(t)}{\rm exp}[(k-1)\frac{n}{\bar{n}(t)}], &n>0
\end{aligned}
\right.
\end{equation}
where $k$ is the ratio $\eta/\alpha$. This equation is obtained under the assumption that $\bar{n}(t)$ is sufficiently large at time $t$. There is a probability when all the electrons are attached, which is the case when $n=0$ in Eq.\ref{eq:ampliexp}, but its probability $k$ can be very small for MRPCs working at high electric field. In the case when $k=0$ which can be regarded as an approximation of the working condition of the MRPC detector, the probability distribution function (PDF) for $n$ at time $t$ becomes:
\begin{equation}
\label{eq:ampliexpsimple}
P(n,t)=\frac{1}{\bar{n}(t)}{\rm exp}[-\frac{n}{\bar{n}(t)}]
\end{equation}

This reveals that the number of electrons $n$ after some specific drifting time also follows an exponential distribution with an average value of $\bar{n}(t)$. This uncertainty purely comes from the avalanche. If $n$ is split into 2 parts $A$ and $\bar{n}(t)$:
\begin{equation}
\label{eq:nsplit}
n(t)=A\bar{n}(t)
\end{equation}
where $\bar{n}(t)$ is the average number of electrons and its expression is shown in Eq.\ref{eq:townsend}. $A$ is a random variable that takes in the uncertainty of the avalanche and follows the exponential distribution: 
\begin{equation}
	\label{eq:Aexpon}
	f(A)=\frac{1}{A_{av}}e^{-\frac{A}{A_{av}}}
\end{equation}
The average value $A_{av}$ should be 1 according to Eq.\ref{eq:ampliexpsimple} and Eq.\ref{eq:nsplit}. Therefore the number of electrons developed from $m$ primary electrons is:
\begin{equation}
\begin{aligned}
	\label{eq:chargewitht}
	N(t)=\sum_{j=1}^{m}n_j(t)=&[\sum_{j=1}^{m}A_j]\bar{n}(t)=Be^{x}\\
	B=\sum_{j=1}^{m}A&_j,\; x=\alpha'vt 
\end{aligned}	
\end{equation}
where $n_j(t)$ is the electrons number as a function of time for the $j$th primary electron. It is formulated as the product of the amplitude $A_j$ and the average number $\bar{n}(t)$, as described by Eq.\ref{eq:nsplit}. The amplitude variable $B$ is the sum of $A_j$ of every single primary electron, and hence, the PDF of $B$ can be obtained using $f(A_j)$:
\begin{equation}
\begin{aligned}
	\label{eq:Bpdf}
	f(B)=f(A_1)\otimes f(A_2)...\otimes f(A_m)=\frac{B^{m-1}e^{-\frac{B}{A_{av}}}}{(m-1)!A_{av}^m}
	\end{aligned}
\end{equation}

The symbol $\otimes$ denotes the {\bf Fourier convolution}, which can be described as:
\begin{equation}
	\label{eq:fourier}
	f(x)\otimes g(x)=\int f(x-\tau)g(\tau)d\tau
\end{equation}

If a fixed threshold $B_{th}$ which is in unit of number of electrons is set to Eq.\ref{eq:chargewitht}, then $x$, that is related with the threshold crossing time $t$ is:
\begin{equation}
	\label{eq:25}
	\begin{aligned}
	x={\rm ln}(\frac{B_{th}}{B}),\;\;\;B=B_{th}e^{-x}
	\end{aligned}
\end{equation}  
Then the PDF of $x$ can be obtained with $f(B)$ and $B(x)$ (Eq.\ref{eq:25}):
\begin{equation}
	\label{eq:pdfx}
	\begin{aligned}
		g(x)=f(B(x))|\frac{{\rm d}B}{{\rm d}x}|=\frac{B_{th}^m}{(m-1)!}{\rm exp}[-B_{th}e^{-x}-mx]	
		\end{aligned}
\end{equation}

\begin{figure}
    \centering
    \begin{subfigure}[b]{0.49\textwidth}
        \includegraphics[width=\textwidth]{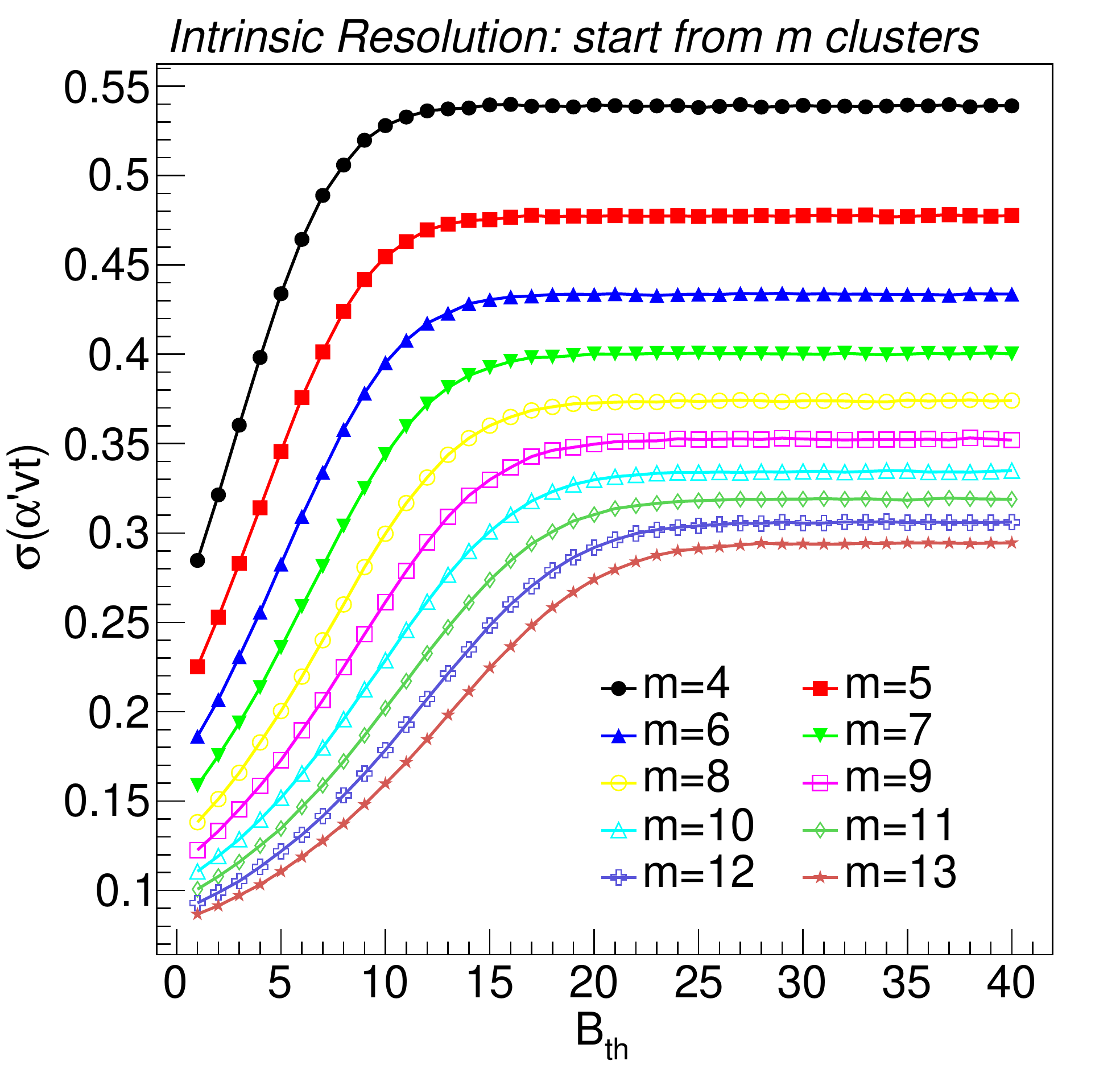}
        \caption{}
        \label{fig:theoresowiththre}
    \end{subfigure}
    \begin{subfigure}[b]{0.49\textwidth}
        \includegraphics[width=\textwidth]{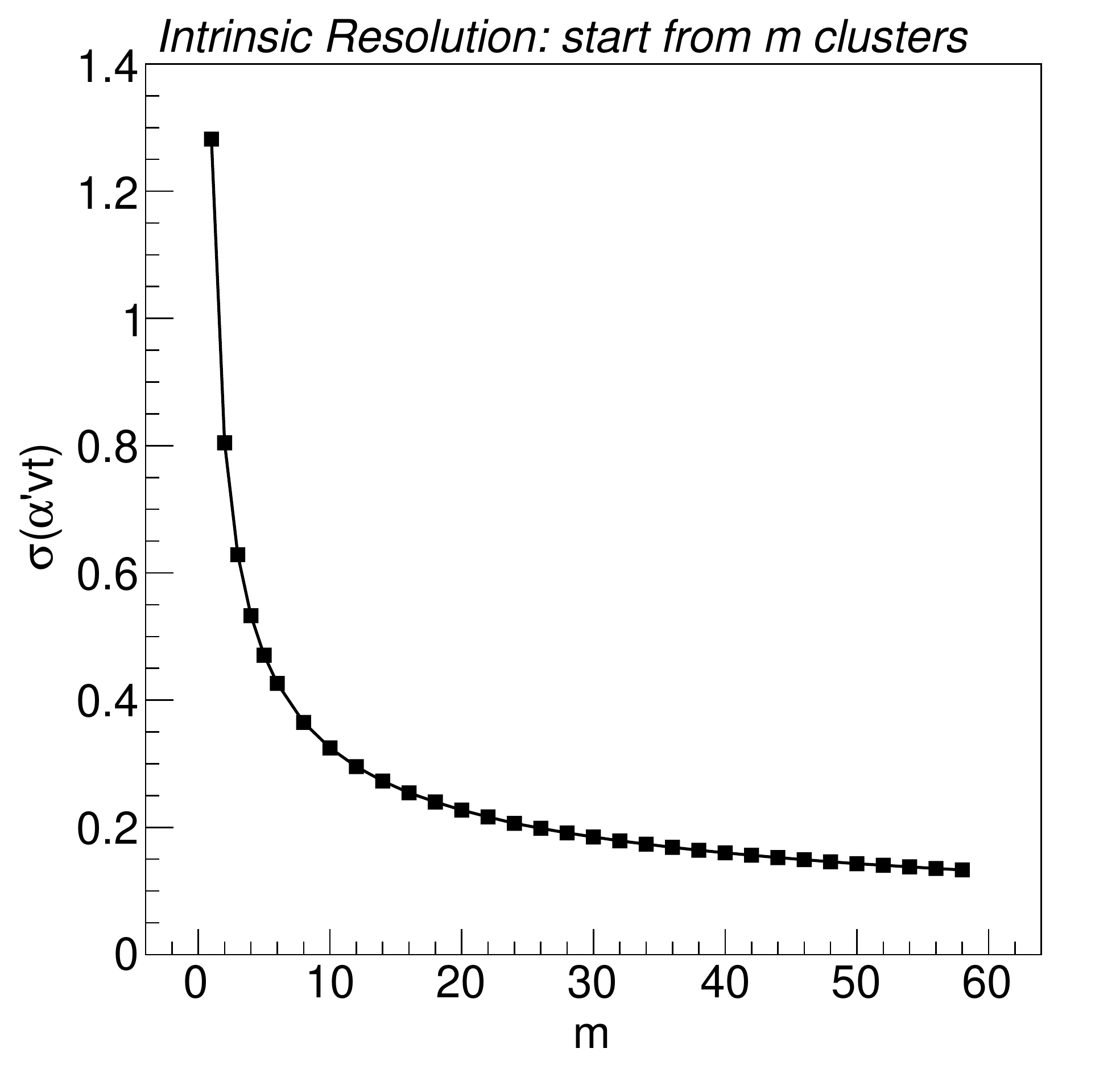}
        \caption{}
        \label{fig:theoresowithm}
    \end{subfigure}
    \caption{(a) The intrinsic time resolution of MRPC detectors with respect to the threshold. Different curves in this plot represent the signals starting from different number of electrons $m$. (b) The intrinsic time resolution with respect to $m$ when $B_{th}=500$.}
    \label{fig:theoreso}
\end{figure}

This equation corresponds to the time response function for $m$ primary electrons when $k=0$ in Ref.\cite{riegler2009time}, but derived in a different way. The standard deviation of $g(x)$ is the intrinsic time resolution in unit of $1/\alpha'v$ for signals starting from $m$ primary electrons. Fig.\ref{fig:theoreso} shows this resolution with respect to the threshold and different number of primary electrons $m$. The intrinsic time resolution is independent of the threshold $B_{th}$, which attributes to the characteristics of the function $g(x)$, where scaling $B_{th}$ is just equivalent to shifting the distribution without changing the standard deviation. This is only true under the assumption $\bar{n}(t)$ is sufficiently large, and that is why the independence relationship appears only when $B_{th}$ is large in Fig.\ref{fig:theoresowiththre}. As defined earlier in this section, $\bar{n}(t)$ is the average electron number developed from one single primary electron. Therefore, in order to ensure that $\bar{n}(t)$ of all the $m$ primary electrons meets this requirement, the criterion for $B_{th}$ to be ``large enough'' increases with $m$. Fig.\ref{fig:theoresowithm} illustrates the relationship between the intrinsic time resolution and $m$ when $B_{th}=500$ which is in the plateau region of Fig.\ref{fig:theoresowiththre} for the $m$ considered here. The time resolution improves with $m$, because the superposition of multiple avalanche averages the signals and reduces its relative standard deviation, making the signal more ``uniform''.

However, in the real MRPC detectors, $m$ is not a constant but itself a random variable. The uncertainty of $m$ mainly comes from the primary deposited energy and the position distribution of the ionizing collisions. It is really hard to summarize an analytical expression of the probability distribution function of $m$, let alone the combination of $m$ and $x$. In this situation, the Monte Carlo simulation of the MRPCs is an effective way and it is described in Sec.\ref{sec:source}.

\section{Different sources that contribute to the intrinsic time resolution} 
\label{sec:source}
To achieve the goal of an intrinsic time resolution below 15 $\rm ps$, it is important not only to know the values of the intrinsic limitation, but also the sources it comes from, because this would help us to better design the MRPC structure. Since only the intrinsic resolution of the detector is considered in this paper, the time jitter brought by the read out systems including the transmission lines and the electronics is neglected. So according to Eq.\ref{eq:pdfx} and its characteristics, the time resolution of the MRPC is attributed to 3 main origins:
\begin{enumerate}
	\item uncertainty of the position where the primary interactions take place,
	\item uncertainty of the deposited energy and the number of primary electrons,
	\item uncertainty of the avalanche multiplication.   
\end{enumerate}
These are studied by Monte Carlo simulations of the detector, which use a simulation framework developed by our group\cite{wang2018simulation}. This framework consists of several modules, each dealing with one relatively independent part of the simulation. For this reason, it is very easy to separate and estimate the contribution of each source.

\begin{table}
\centering
\small
\begin{tabular}{| c | c | c | c | c | }
\hline
 Experiments & StackNo. & GapNo. & Thickness[$\rm \mu m$] & Working E [kV/cm] \\
\hline
ALICE & 2 & 5 & 250 & 104 \\
\hline
CBM & 2 & 4 & 250 & 110 \\
\hline
STAR & 1 & 6 & 220 & 114 \\
\hline
BESIII & 2 & 6 & 220 & 103 \\
\hline
RefMRPC & 4 & 6 & 160 & 135 \\
\hline
THU1 & 4 & 8 & 104 & 159 \\
\hline
THU2 & 1 & 6 & 250 & 109 \\
\hline
\end{tabular}
\caption{The inner geometry and working condition of several typical MRPC detectors.}
\label{tab:geooftypicalMRPC}
\end{table}

\begin{figure*}[h!]
	\centering
	\includegraphics[width=0.9\textwidth]{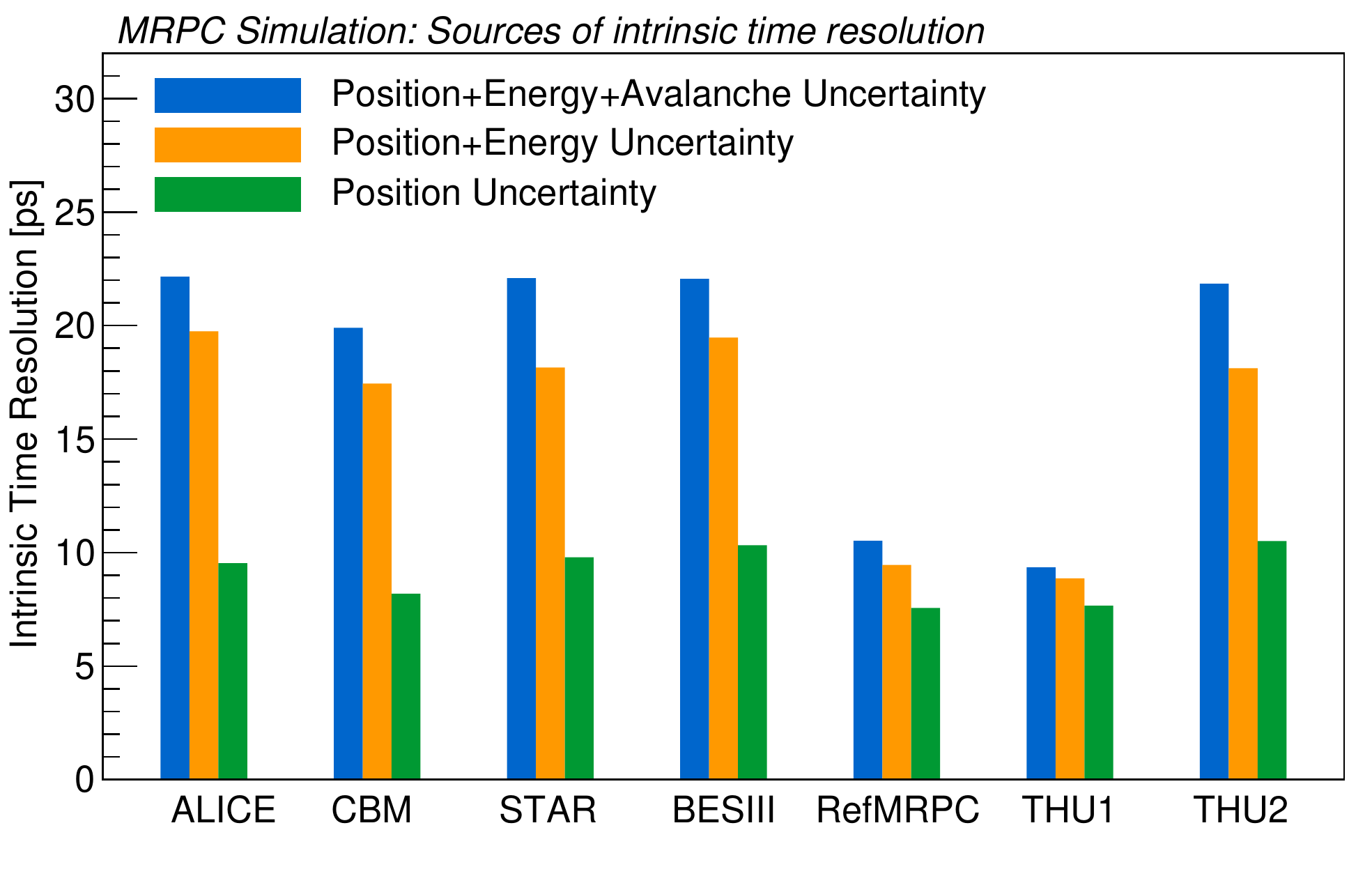}
	\caption{The contributions to the intrinsic time resolution from different sources for the MRPCs presented in Tab.\ref{tab:geooftypicalMRPC}.}
	\label{fig:sourceshist}
\end{figure*}

In this section, some typical MRPCs\cite{akindinov2009final,Toia:209729,bonner2003single,yong2012prototype} that are or will be used in several high energy physics experiments including ALICE\cite{alice2014performance}, CBM\cite{ablyazimov2017challenges}, STAR\cite{ackermann2003star} and BESIII\cite{besiii2009construction} are explored. RefMRPC, a very thin-gap detector\cite{an200820} and 2 prototypes THU1 and THU2 assembled in our laboratory\cite{wang2018sorma} are also included. Tab.\ref{tab:geooftypicalMRPC} shows the inner geometry and the working electric field of these detectors and Fig.\ref{fig:sourceshist} shows the quantitative contributions from the sources mentioned above. The electric field in the gas gap is set with the value shown in Tab.\ref{tab:geooftypicalMRPC}, which is the same as the working field given in the references of the corresponding experiments. Standard gas mixture of 90\% $\rm C_2H_2F_4$, 5\% $\rm SF_6$ and 5\% $\rm iso$-$\rm C_4H_{10}$ is used in all the simulations in this paper. 

\begin{figure*}[h!]
	\centering
	\includegraphics[width=0.85\textwidth]{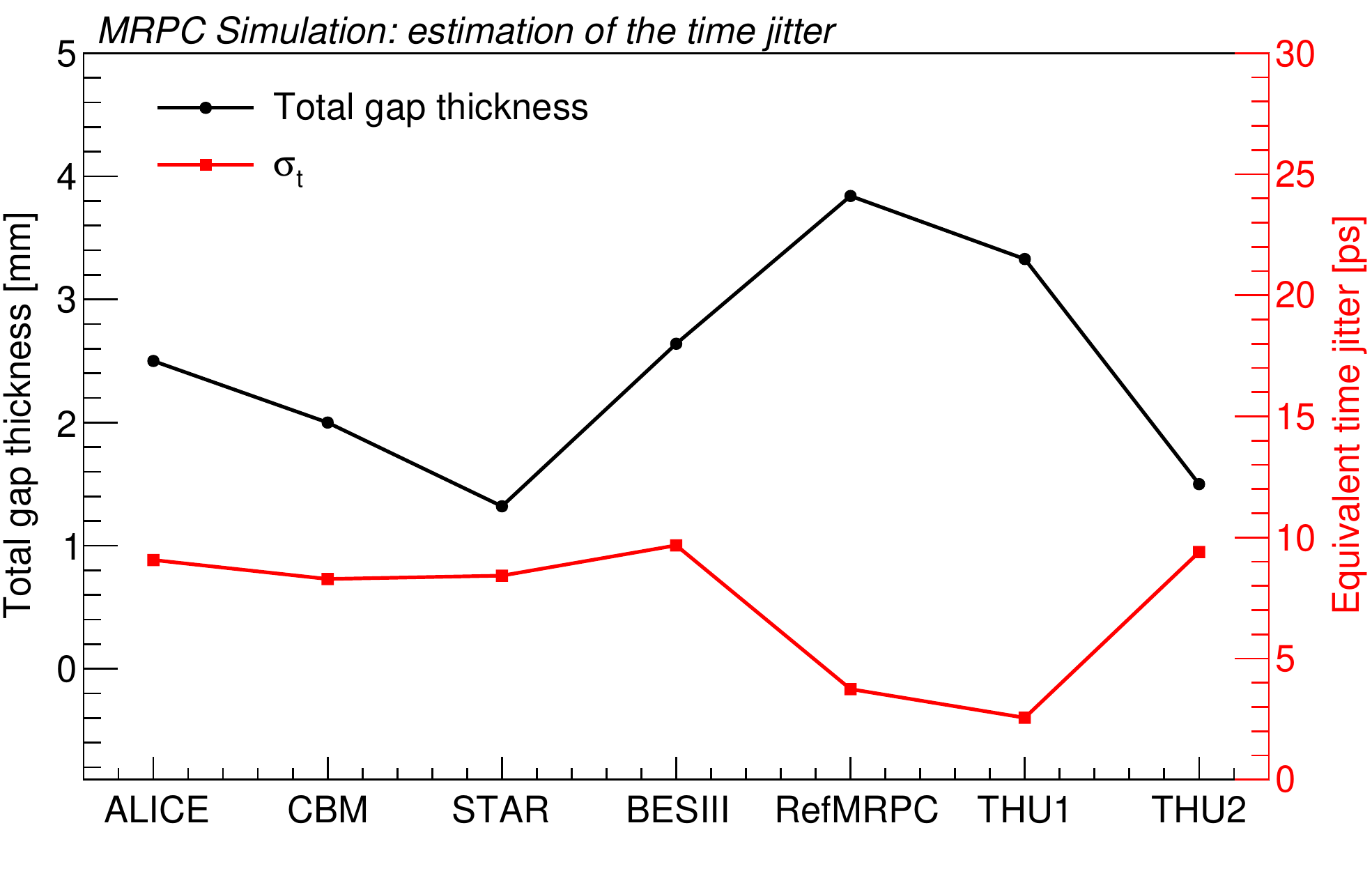}
	\caption{The total gap thickness and the equivalent time jitter for the MRPCs presented in Tab.\ref{tab:geooftypicalMRPC}.}
	\label{fig:estimatesigmat}
\end{figure*}

In Fig.\ref{fig:sourceshist}, ``Position Uncertainty'' includes only the uncertainty of the interaction position, while the total deposited energy for every event is kept to be its average value in the full simulation, and the avalanches of the ionized electrons strictly obey the exponential growth. In this situation, the intrinsic time resolution of all the MRPCs in Tab.\ref{tab:geooftypicalMRPC} is on the order of 10 $\rm ps$, and does not vary much among different detectors. The slight differences should come from the statistical errors, the electric field and the total thickness of the detector. As only the uncertainty of the position is considered, the number of electrons as a function of time in these detectors is:
\begin{equation}
	\label{eq:Nandm}
	\begin{aligned}
	N(t)=me^{\alpha'vt}
	\end{aligned}
\end{equation}
where $m$ is the number of primary electrons that develop into effective avalanche ($e^{\alpha'vt}\gg1$) and is the only random variable in the formula. At the threshold crossing time, the slope of the signal is $s=dN/dt=\alpha'v\cdot me^{\alpha'vt}$, and the equivalent time jitter is:
\begin{equation}
	\label{eq:estimatet}
	\begin{aligned}
	\sigma_t=\frac{\sigma_{N(t)}}{s}=\frac{\sigma_me^{\alpha'vt}}{\alpha'v\cdot me^{\alpha'vt}}=\frac{1}{\alpha'v}\cdot\frac{\sigma_m}{m}
	\end{aligned}
\end{equation}
where $\sigma_m$ is the standard deviation of $m$. Eq.\ref{eq:estimatet} demonstrates that the intrinsic time resolution from only the uncertainty of the interaction position is affected by the effective Townsend coefficient $\alpha'$, the drift velocity $v$ and the relative standard deviation of $m$. The former 2 coefficients are determined by the electric field, while $\sigma_m/m$ is negatively corrected with the total gap thickness. Fig.\ref{fig:estimatesigmat} shows the $\sigma_t$ and the total gap thickness for all the MRPCs presented in Tab.\ref{tab:geooftypicalMRPC} with the red and black markers respectively. The relationship of $\sigma_t$ is consistent with the green bars in Fig.\ref{fig:sourceshist}. A larger $\alpha'v$ at higher electric field leads to a much more intense avalanche multiplication, making the signal grow faster and thus less affected by $\sigma_m/m$. RefMRPC and THU1 have both large working electric field and total gap thickness, so their intrinsic time resolution is the smallest, while the performances of the BESIII and THU2 MRPCs are restricted by the electric field and the thickness respectively. As for the STAR and CBM MRPCs, although the electric field of STAR is higher, its resolution is still worse than the CBM and this is resulted from the smaller total gap thickness.

\begin{figure*}[h!]
	\centering
	\includegraphics[width=0.56\textwidth]{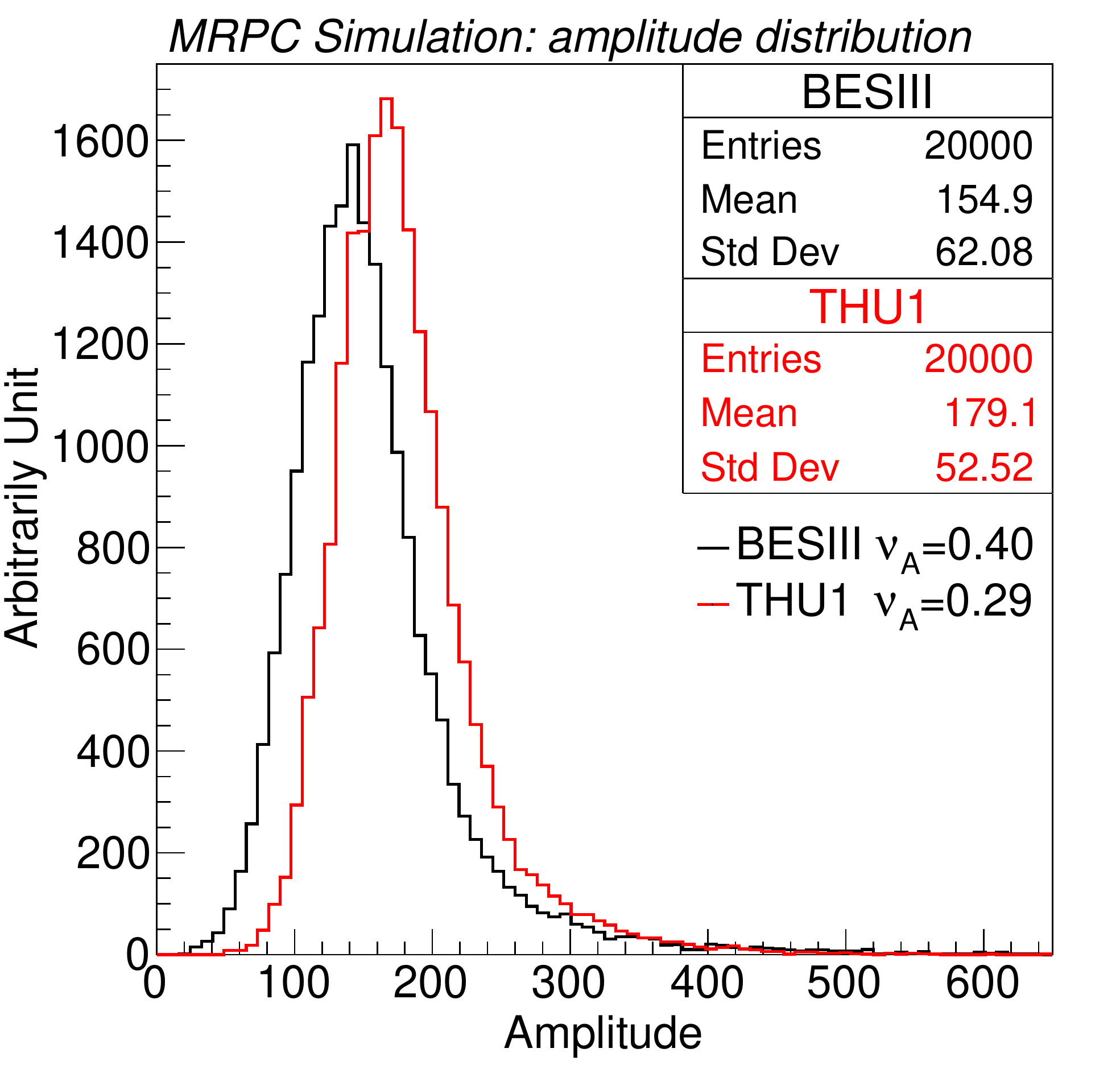}
	\caption{The distribution of the signal amplitude for BESIII and THU1 MRPC detectors.}
	\label{fig:amplidist}
\end{figure*}

``Position+Energy Uncertainty'' in Fig.\ref{fig:sourceshist} shows the results when the uncertainty from the deposited energy is added to ``Position Uncertainty''. In this case, the number of the primary electrons has a much larger variance, making some of the ionized clusters to contain large amounts of electrons (large clusters), while some only a few (small clusters). This worsens the time resolution. It can be clearly seen from Fig.\ref{fig:sourceshist} that this effect is obvious in thick-gap detectors, but not so obvious in thin-gap ones. It is essentially due to the electric field, which is relatively low in thick-gap MRPCs in order to avoid streamers. The reason is similar to the ``Position Uncertainty''. In the condition of low electric field, the uncertainty for the small clusters to be effective is very large, with a strong dependence on the position where they are generated in the gas gap. The multiplication of large clusters in the same event will finally be restricted by the space charge effects\cite{riegler2003detector} and will not be able to compensate for the loss of the small ones. Thus, the variation of the signal amplitude turns to be relatively large, with consequences on the dynamic range of the time resolution. However, when the electric field becomes higher, the possibility for the clusters to become effective increases, especially for small clusters. The variation of the amplitude is hardly affected by the energy deposition fluctuations. Fig.\ref{fig:amplidist} shows the distribution of the signal amplitude for BESIII and THU1 MRPC detectors which have the lowest and the highest electric field in Tab.\ref{tab:geooftypicalMRPC}. The relative standard deviation of the amplitude $\nu_A$ for THU1 is 0.29, while that for BESIII is about 0.4. That is why RefMRPC and THU1 has a much better time resolution than the others.

``Position+Energy+Avalanche Uncertainty'' considers also the variance of the size of the avalanche. The time resolution brought by the avalanche can be explained by Fig.\ref{fig:theoreso}. When the total gap thickness of the detector is large, the incident particles will have a longer track and thus a larger $m$, which indicates a better time resolution as shown in Fig.\ref{fig:theoresowithm}. So the small total gap thickness is the main reason for STAR and THU2 detectors to have a relatively large variance of avalanche. It is also obvious that the uncertainty brought by the avalanche is much smaller for RefMRPC and THU1 compared to other detectors. On the one hand, this is the result of the extremely large total gap thickness in these two detectors, and on the other hand, the high electric field guarantees a larger effective Townsend coefficient $\alpha'$, and thus an improved time resolution ($\sigma(t)=\sigma(x)/\alpha'v$, where $\sigma(t)$ is the time resolution and $x=\alpha'vt$ as defined in Sec.\ref{sec:theory}). The results shown in Fig.\ref{fig:sourceshist} are consistent with the conclusions in Sec.\ref{sec:theory}, but obviously, the uncertainty of the avalanche is small and does not vary much from detectors to detectors. This is because in the gas condition of the simulation, the average number of clusters per millimeter is around 10 /mm, and the average number of electrons for every cluster is around 2.7. Though $m$ is uncertain for all the MRPCs in Tab.\ref{tab:geooftypicalMRPC}, considering the total gap thickness, it is absolutely a number larger than $15$. In Fig.\ref{fig:theoresowithm}, this is the value where the resolution is already small and decreases slower.

In general, large total gap thickness and small single gap thickness are two key factors that bring a good time resolution, because they guarantee a large number of primary electrons $m$ and high electric field. This fundamental knowledge about the intrinsic resolution is important not only for understanding the physics of the MRPC, but also the design and construction of the high resolution detectors. Therefore they are quantitatively studied in Sec.\ref{sec:geo} together with some other parameters of the inner geometry of the detector. 

\section{The impact of the MRPC inner geometry on the intrinsic time resolution} 
\label{sec:geo}
The structure of the detector geometry has a significant impact on MRPC's performance. As mentioned above, MRPCs used in many present and future physics experiments have a time resolution around 60 $\rm ps$. This is because the gap thickness is all around 220$\sim$250 $\rm \mu m$. According to Fig.\ref{fig:sourceshist}, the 20 $\rm ps$ timing requirement for the MRPC system of the SoLID experiment makes the design with thin gaps to be the only promising choice. This section presents the quantitative results of the intrinsic time resolution of MRPC with different inner geometry, aiming to provide guidance for designing the detectors.

\begin{figure*}[h!]
	\centering
	\includegraphics[width=0.56\textwidth]{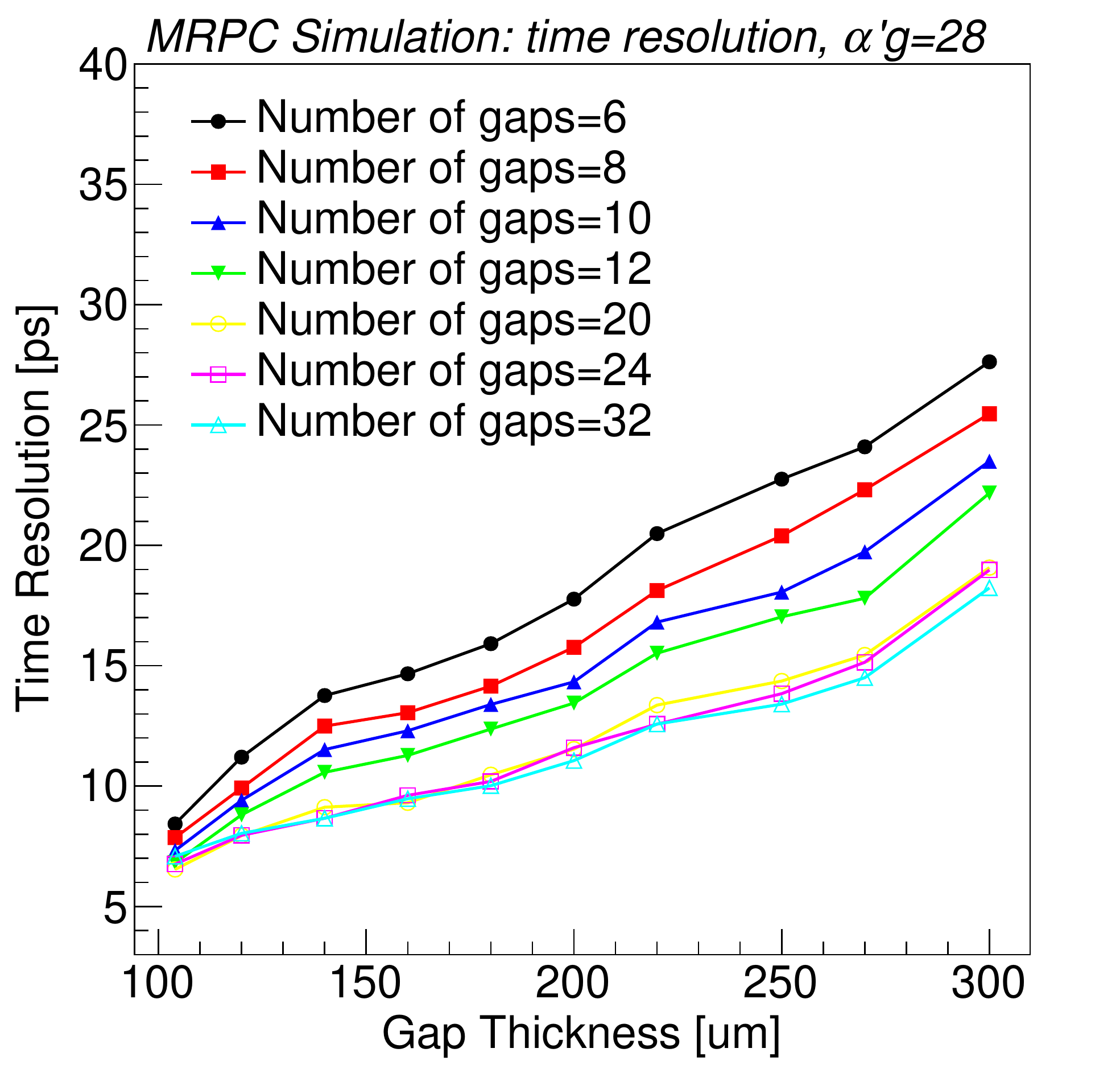}
	\caption{The time resolution of MRPCs with different thickness and numbers of the gas gaps.}
	\label{fig:geowiththick}
\end{figure*}

\begin{figure}
    \centering
    \begin{subfigure}[b]{0.49\textwidth}
        \includegraphics[width=\textwidth]{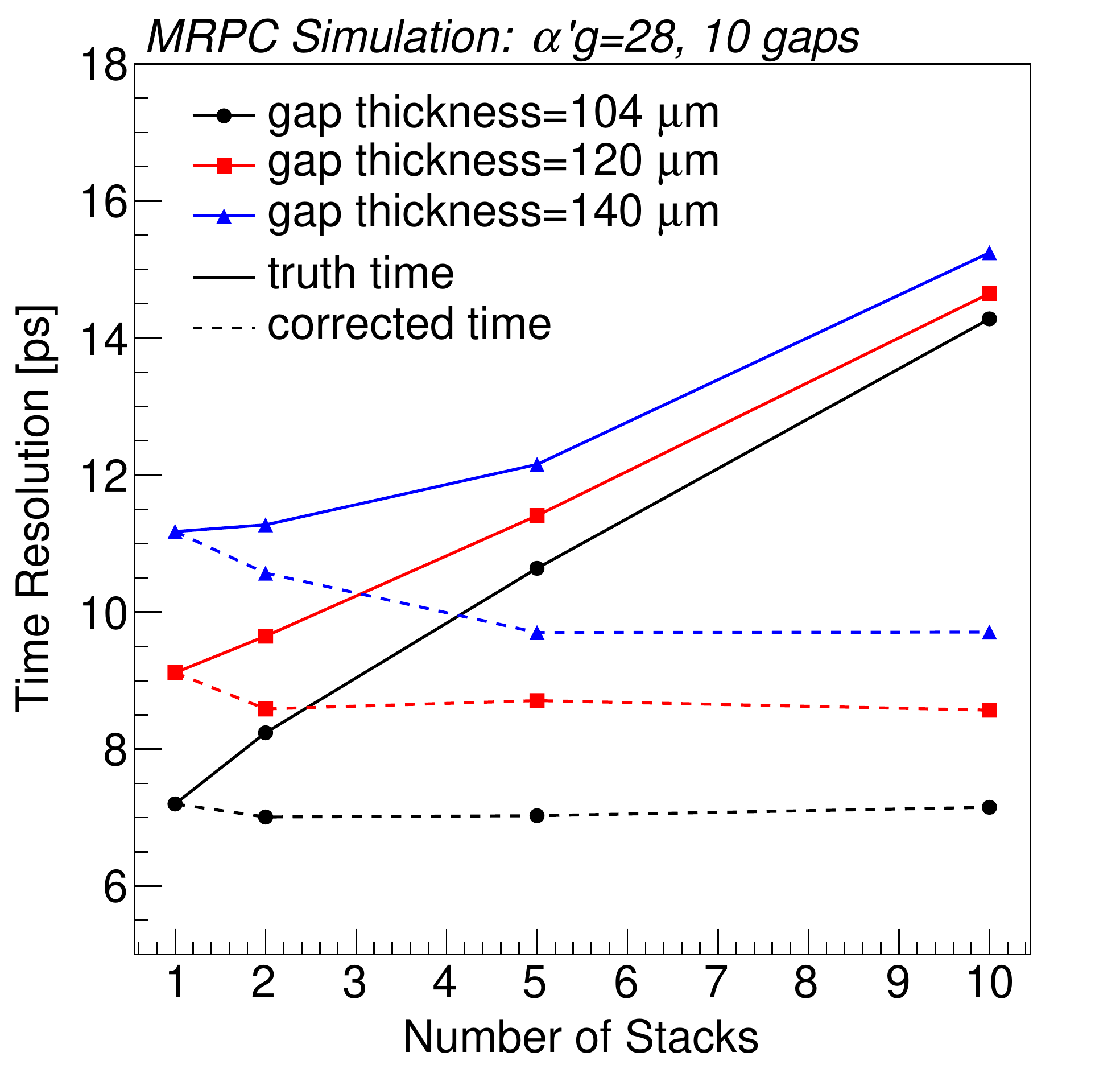}
        \caption{}
        \label{fig:geowithstack10}
    \end{subfigure}
    \begin{subfigure}[b]{0.49\textwidth}
        \includegraphics[width=\textwidth]{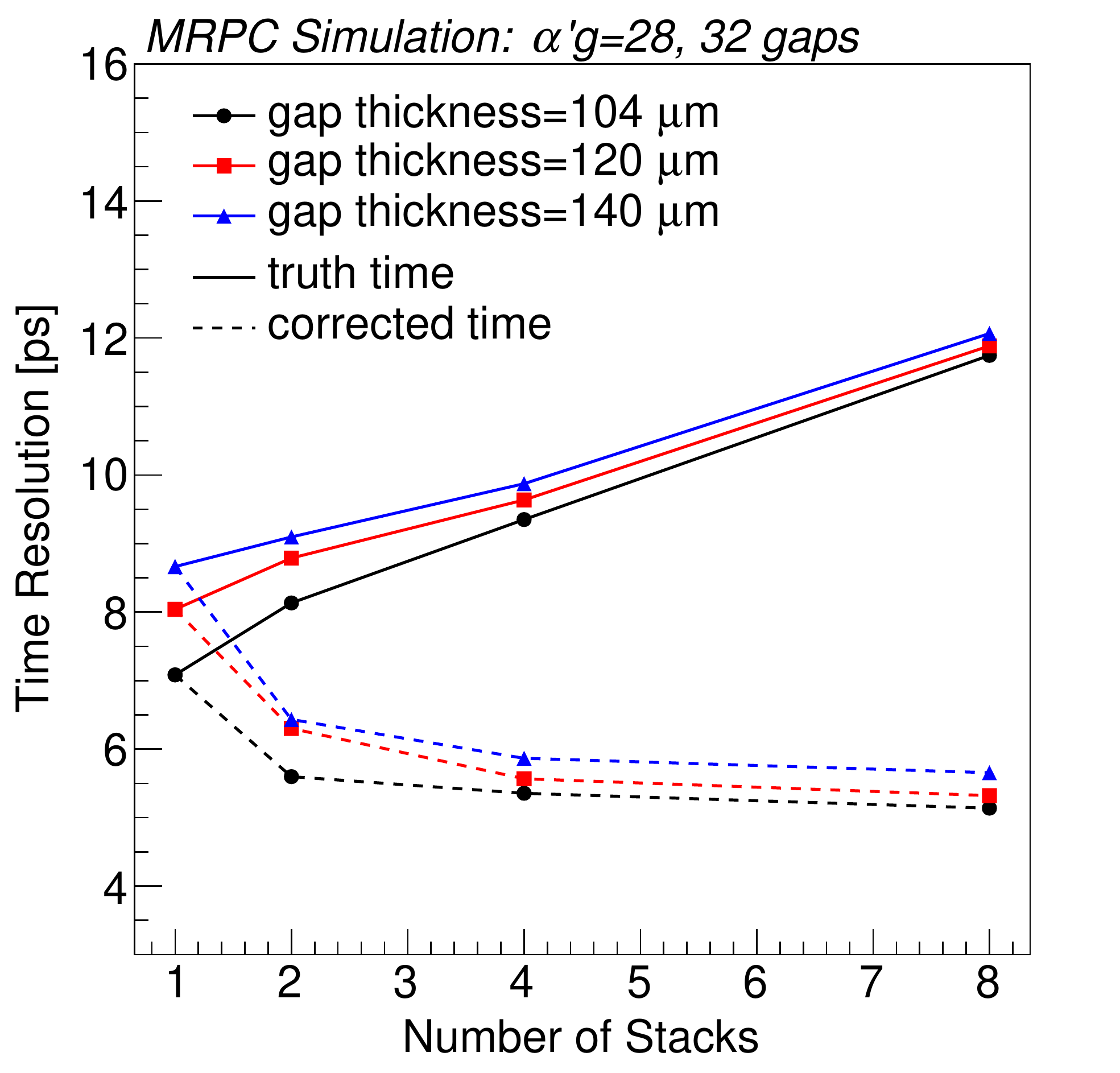}
        \caption{}
        \label{fig:geowithstack32}
    \end{subfigure}
    \caption{The time resolution of MRPCs with different number of stacks in a detector. The total number of gaps are fixed to be 10 in (a) and 32 in (b).}
    \label{fig:geowithstack}
\end{figure}

Fig.\ref{fig:geowiththick} shows the time resolution with respect to the single gap thickness and the number of gaps, while Fig.\ref{fig:geowithstack} presents the resolution with different number of stacks. To control the final avalanche size and avoid streamers, the effective Townsend coefficient times the gap thickness for all the points in Fig.\ref{fig:geowiththick} and Fig.\ref{fig:geowithstack} is fixed at $\alpha'g=28$, which is close to a normal working condition of the MRPCs shown in Sec.\ref{sec:source}. Thinner and more gaps clearly have a better time resolution. Another interesting point is that when the number of gaps goes from 6 to 12, the time resolution drops significantly, but it remains almost unchanged when the number of gaps is over 20. This corresponds to Fig.\ref{fig:theoresowithm}, where the derivative of $\sigma$ with respect to $m$ is a decreasing function. Based on Fig.\ref{fig:geowiththick} and considering the goal of 15 $\rm ps$ for only the detector, the possible choices for next generation MRPCs should be the designs with thickness below 160 $\rm \mu m$. It is worth noting that the intrinsic time resolution with respect to the number of gaps is also studied in Ref.\cite{riegler2003detector} but only for gap thickness $g=300$ $\rm\mu m$ and electric field $E=100$ $\rm kV/cm$. The intrinsic time resolution of $g=300$ $\rm \mu m$ MRPC simulated in this paper is consistent with the reference, while the tiny difference may come from the differences between the gas mixture.

The curves with larger number of gaps give a much better result than smaller ones as is presented in Fig.\ref{fig:geowiththick}. However in actual cases, the voltage needed by these detectors is much higher if all the gaps are in one single stack, which significantly increases the cost. Usually, gaps are divided into several stacks and the voltage is added separately on each stack. However, when the detector is divided into more stacks, it will become much thicker, due to the additional PCB layers for the high voltage electrodes and at least one more glass plate per stack. This will lead to a longer time interval between the very first ionized electron in the top gap and the very last in the bottom. Thus the variance of the starting time for the avalanches in the whole detector becomes larger, which results in a larger time resolution. This is shown in Fig.\ref{fig:geowithstack} by the solid curves. The total number of gas gaps is fixed to be 10 in Fig.\ref{fig:geowithstack10} and 32 in Fig.\ref{fig:geowithstack32}. The dashed curves are the time resolution if the arrival time of the incident particle at each stack is corrected to be the same. In this ``corrected'' case, the resolution improves with respect to the number of stacks, because the starting time of each avalanche is limited inside only one stack, and the thickness of a single stack will become thinner when the number is increased. The improvement of the time resolution is more obvious in the case of more gaps, as a result of a much larger decrease of the stack thickness. The rules are the same for three kinds of the gap thickness in the figure and so are all the other choices of the thickness. 

\section{The intrinsic time resolution analyzed with neural network} 
\label{sec:NN}
In the previous sections, the intrinsic time resolution either obtained by theoretical derivations or by Monte Carlo simulations has the same assumption --- the time reconstruction algorithm is the so-called ToT method. This is a traditional algorithm and used nearly in all the present experiments. However, if this assumption is changed and another algorithm is applied, the intrinsic time resolution could be completely different. This section compares the results from ToT and the neural networks.

\begin{figure*}[h!]
	\centering
	\includegraphics[width=0.56\textwidth]{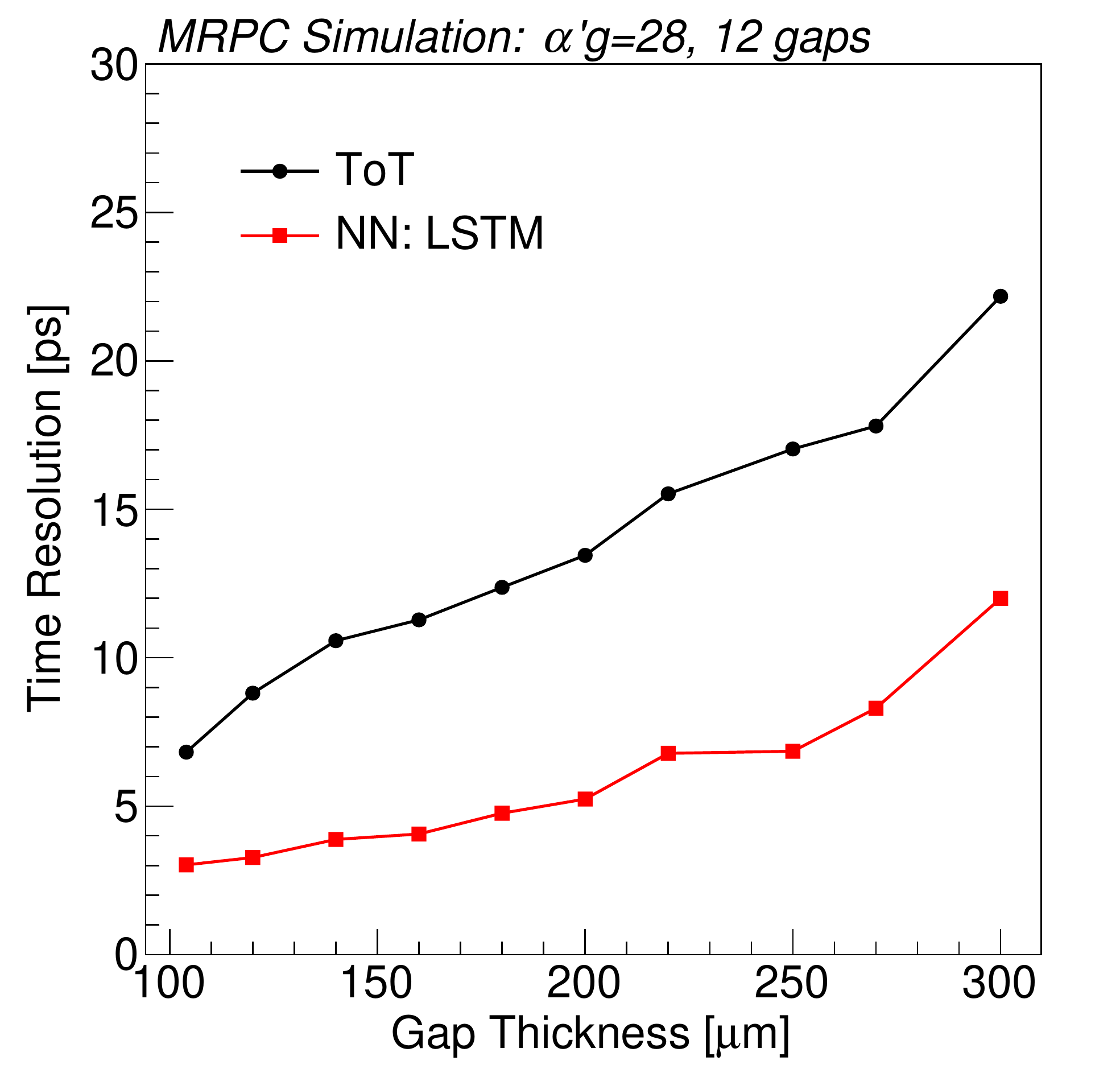}
	\caption{The comparison of the intrinsic time resolution analyzed with two different algorithms.}
	\label{fig:resocompare}
\end{figure*}

MRPCs with 12 gas gaps and different gap thickness are simulated and analyzed. Same as Sec.\ref{sec:geo}, the $\alpha'g$ is fixed at 28. Fig.\ref{fig:resocompare} shows the results obtained with the ToT and Long Short Term Memory network (LSTM)\cite{lecun2015deep}. LSTM is a special case of the Recurrent Neural Network (RNN)\cite{lecun2015deep} which has been widely used in many temporal problems. By analyzing all the data of the waveform, LSTM finds a very good connection between the signal and the particles arrival time. The intrinsic time resolution becomes better. Details of implementation and the main idea of this network can be found in Ref.\cite{wang2018sorma}. The result is reasonable, because algorithmically, ToT is just one specific model which is designed by the scientists and is proved to be useful. But neural networks, no matter what network specifically, can be recognized as a set of models because of its huge number of parameters and the complexity of the structure. The goal of the network training is merely to find a model in the model set that solves the problem best, which means this new algorithm is more generic and thus more powerful in formulating problems. 

\section{Conclusions}
\label{sec:concl}
A detailed study of the intrinsic time resolution for future MRPC detectors is presented in this paper. Theoretical derivations of the intrinsic time resolution for MRPC signals are described. Based on the theory and a Monte Carlo simulation, the sources of the timing uncertainty and the effects brought by the inner geometry of the detector are carefully analyzed. Finally, a comparison of the intrinsic resolution obtained with another reconstruction algorithm --- the neural network is made and the obtained results are better than with the traditional ToT algorithm. This paper proves that the goal of the 20 $\rm ps$ MRPC system is promising from the intrinsic point of view, and provides useful guidance about how to design the real detector. Meanwhile, the intrinsic time resolution summarized in this work also provides useful results for all the other groups working on MRPC detectors.

\section{Acknowledgments}
The work is supported by National Natural Science Foundation of China under Grant No.11420101004, 11461141011, 11275108, 11735009. This work is also supported by the Ministry of Science and Technology under Grant No. 2015CB856905, 2016 YFA0400100.

\bibliographystyle{elsarticle-num}
\bibliography{myrefs.bib}{}
%
%

\end{document}